# Dissipationless Anomalous Hall Current in $Fe_{100-x}(SiO_2)_x$ Films

W. J. Xu[1], B. Zhang[2], Z. Wang[1], W. Li[1], R. H. Yu[3],

and X. X. Zhang[2*]

[1]Dept. of Phys. and Institute of Nanoscience & Technology, The Hong Kong University of Science and Technology (HKUST), Clear Water Bay, Kowloon, Hong Kong, P. R. China

[2]Image-characterization Core Lab, Research and Development, 4700 King Abdullah University of Science and Technology (KAUST), Thuwal 23900-6900, Kingdom of Saudi Arabia

[3]School of Materials Science and Engineering, Beihang University, Beijing 100191, P. R. China

**Abstract**

The observation of dissipationless anomalous Hall current is one of the experimental evidences to confirm the intrinsic origin of anomalous Hall effect. To study the origin of anomalous Hall effect in iron, $Fe_{100-x}(SiO_2)_x$ granular films with volume fraction of $SiO_2$ $0 \leq x \leq 40.51$ were fabricated using co-sputtering. Hall and longitudinal resistivities were measured in the temperature range 5 to 350 K with magnetic fields up to 5 Tesla. As x increased from 0 to 40.51, the anomalous Hall resistivity and longitudinal resistivity increased about 4 and 3 orders in magnitude, respectively. Analysis of the results revealed that the normalized anomalous Hall conductivity is a constant for all the samples, the evidence of dissipationless anomalous Hall current in Fe.

* Xixiang.Zhang@kaust.edu.sa

One of the most important experimental studies about the intrinsic mechanism of anomalous Hall effect (AHE) in ferromagnetic materials is the observation of dissipationless anomalous Hall current [1]. Although the dissipationless feature of the anomalous Hall conductivity, AHC ( $\sigma_{xy}^{A}$ ), was predicted theoretically long time ago [2, 3], it was observed only recently in the ferromagnetic spinel $CuCr_2\ Se_{4-x}Br_x$ by Lee *et al.* [1]. As pointed in Ref. 1, the direct test of the dissipationless nature of Hall current is to examine whether the anomalous Hall conductivity $\sigma_{xy}^{A}$ (defined as $\rho_{xy}^{A}/\rho_{xx}^{2}$ ) changes with increasing the impurity or with decreasing the relaxation time of electrons. In other words, the demonstration of the dissipationless anomalous Hall current is the observation of the square dependence of the normalized Hall resistivity (by the carrier density) on the longitudinal resistivity, i.e., $\rho_{xy}^{A}/n \propto \rho_{xx}^{2}$ . In Ref. 1, the Hall resistivity and longitudinal resistivity increased about 5 and 3 orders in magnitude respectively as x varied from 0 to 1, while keeping the ferromagnetic state nearly the same. More than three-order changes in both qualities indeed guaranteed a reliable fitting. In most of the studies, the experimental data were fitted to the scaling law $\rho_{xy}^{A} \sim \rho_{xx}^{\gamma}$ . Since the changes in longitudinal resistivity and AHE resistivity were often too narrow to assure a reliable linear fitting in a log-log plot, the fitted values of $\gamma$ were varied from 0.7 to 3.7 in different systems [4-9]. In the models of extrinsic anomalous Hall effect, $\gamma$ =1 and $\gamma$ =2 indicate the skew scattering and side-jump mechanisms, respectively [10,11].

The intrinsic nature of AHE in iron has been theoretically demonstrated by Yao *et al.* in 2004 [12]. Recently AHE in Fe has been studied by Tian *et al.* using epitaxial films with different thicknesses [13]. To separate the intrinsic and extrinsic contributions from the measured AHE, they proposed that the residual resistivity should be involved in the scaling. AHE in Co has been experimentally studied using films with different thicknesses by Kotzler and Gil [14]. A temperature independent anomalous Hall conductivity was observed, i.e., $\rho_{xy}^{A}(T) \propto \rho_{xx}^{2}(T)$, where $\rho_{xy}^{A}(T)$ and $\rho_{xx}(T)$ are measured from the same sample at different temperature. Actually, this observation is different from that of the dispassionless anomalous Hall current [1].

In this letter, we report the observation of the dissipationless anomalous Hall current in $Fe_{100\text{-}x}(SiO_2)_x$ granular thin films with $SiO_2$ volume fraction $0 \le x \le 40.51$. This should be the first experimental demonstration of the intrinsic anomalous Hall effect in ferromagnetic iron [12].

More than 13 $Fe_{100\text{-}x}(SiO_2)_x$ granular films with $SiO_2$ volume fraction $0 \le x \le 40.51$ were fabricated by co-sputtering on glass and Kapton substrates. The base pressure and argon gas pressure for sputtering were $2\times10^{-7}$ Torr and $4\times10^{-3}$ Torr, respectively. During the deposition the substrates were at room temperature. Different $SiO_2$ volume fractions were obtained by controlling the relative sputtering powers applied to Fe and $SiO_2$ targets and confirmed by X-ray fluorescence (XRF) measurements. The film thickness was controlled by the sputtering time then measured with a Veeco-Dektak surface profiler. To perform Hall and longitudinal resistivity measurements simultaneously and to guarantee that the data for electrical transport (Hall and longitudinal resistivity) measurements were obtained from the same sample, masks were used to make patterned samples**.** Patterned films on glass substrates were used for the resistivity and Hall measurements with a Quantum Design Physical Property

Measurement System. Samples deposited on Kapton substrates were used for magnetic measurements using Quantum Design Magnetic Property Measurement System.

The dissipationless feature of AHC can be demonstrated by the observation of a constant Hall conductivity normalized by carrier density ($\sigma_{xy}^{A}/n$ , where n is the carrier density) with increasing the impurity/imperfection in the ferromagnetic materials [1]. In the present study, the ferromagnetic Fe was doped by $SiO_2$ . As the $\sigma_{xy}^{A}$ can be approximated by $\rho_{xy}^{A}/\rho_{xx}^{2}$ , we can easily fit the data to a power law dependence $\rho_{xy}^{A}/n \sim \rho_{xx}^{\gamma}$ . If the fitted exponent $\gamma$ =2, one may claim that the anomalous Hall conductivity is dissipationless, or the measured anomalous Hall effect is intrinsic [1].

The Hall resistivity, $\rho_{xy}$, for all samples was measured in a magnetic field $-5T \leq H \leq 5T$ and at temperatures ranging from 5 K to 350 K. Fig. 1 shows the field dependence of Hall resistivity of $Fe_{94.64}(SiO_2)_{5.36}$ and $Fe_{59.49}(SiO_2)_{40.51}$ at different temperatures. These data are the total Hall resistivities that include both the ordinary Hall effect (OHE) and anomalous Hall effect (AHE) contributions. Normally, the Hall resistivity in ferromagnetic materials can be described as [15]

$$\rho_{xy} = R_O H + R_S 4\pi M \tag{1}$$

where $\rho_{xy}$ is Hall resistivity. $R_O H$ is the ordinary Hall effect (OHE) caused by the Lorentz force, a classical mechanism. The second term, $R_S 4\pi M$ is the anomalous Hall effect (AHE) and has a quantum mechanics origin [2,3,10,11]. As expected, the field dependent Hall resistivity curves (Fig.1a & b) have very similar behaviors to that of field-dependent magnetization curves with the typical characteristics of a soft ferromagnetic film whose magnetization lies in the film plane. With the increase of a perpendicular magnetic field, the magnetization is forced to rotate to the field direction gradually, which results in a linearly field-dependent magnetization until saturation at about 14 kOe and 11 kOe for x=5.36 and 40.51 respectively. Due to the liner dependence of AHE on magnetization (Eq. 1), a linear change in Hall resistivity is observed at low fields and saturation at high fields.

Above the saturation field, the weak, linear increase of Hall resistivity is due to the ordinary Hall effect, the first term in the right hand of Eq. 1. By fitting this linear part, we obtained the ordinary Hall coefficients ($R_0$) that were used to calculate the effective carrier densities of the samples. At the same time, the saturated AHE was also extracted. Fig. 1c shows the saturated AHE obtained at 5 Kelvin for all the samples with different compositions. It is seen that, in overall, AHE increases as x increases. The ratio of $\rho_{xy}^{A}(x=40.51)/\rho_{xy}^{A}(0)$ is about 14275, *more than 4 orders of increase in magnitude*. The giant increase in AHE is often referred to as a giant Hall effect (GHE) [6]. One will see below that this giant change in AHE is essential to make a reliable scaling between AHE and the longitudinal resistivities.

Shown in Fig.2 are the temperature dependent longitudinal resistivities measured under zero magnetic field from 5 K to 350 K for samples with various compositions. It is evident that the behaviors of the resistivity curves changed as the volume fraction of $SiO_2$ increased. For pure Fe and samples with x smaller than 30, resistivity increased monotonically with temperature, a typical metallic behavior. When x > 30, the behavior of the resistivity changes to a semiconductor-like one, i.e., resistivity decreases monotonically with increasing temperature. In Fig. 2b we plot the 300 K temperature coefficient of the resistivity (TCR), $d(\ln\rho)/d(\ln T)$, as a function of x. A sign change at x ~ 30 is clearly seen. The positive and negative TCR normally indicate a metallic behavior and a semiconductor behavior, respectively. For pure Fe, the temperature dependence of resistivity is mainly governed by phonon scatterings. With the increase of the $SiO_2$ volume fraction, more and bigger insulating clusters formed in the Fe, and finally the whole film become insulating, a typical percolation system [16]. When the $SiO_2$ volume fraction approaches the percolation threshold, the material shows very weak temperature dependence. This is because the phonon scatterings become less important in comparison with the scatterings by the insulating impurities and strong structure disorder. This leads to a decreasing TCR (Fig. 2b). Above the percolation threshold, the electrical transport is dominated by hopping/tunneling processes of electrons. A semiconductor-like behavior appears and TCR becomes negative. Normally, the longitudinal resistivity can increase by a few orders in magnitude at the percolation threshold [1]. In this system, the resistivity (at 5K) increased about 800 times as x increased from 0 to 40.51.

Now let us understand the opposite trends of Hall resistivity with temperature in the two samples shown in Fig. 1a&b. Hall effect/AHE increases with temperature for sample x=5.36; whereas the Hall effect/AHE decreases with temperature for sample with x=40.51. The behavior of AHE resistivity depends strongly on the behavior of longitudinal resistivity, based on the scaling laws for intrinsic (Eq. 2) or extrinsic (Eq. 3) anomalous Hall effect:

$$\rho_{xy}^{A} \big/ n \propto \rho_{xx}^{2} \ , \tag{2}$$

$$\rho_{xy}^{A} \sim \rho_{xx}^{\gamma} \ . \tag{3}$$

For the sample with x=5.36, the longitudinal resistivity increases monotonically with temperature, which leads to a monotonically increase of AHE. However, the resistivity decreases monotonically with temperature for x = 40.51, which certainly leads to a decreasing AHE with temperature.

As discussed previously, by fitting the AHE as a function of magnetic field above the saturation the ordinary Hall coefficient can be extracted. We therefore calculated the effective carrier density using these OHE coefficients ($R_O$) and $n = \frac{1}{R_O e}$, where $e$ is the electronic charge. Shown in Fig. 3 are the calculated effective carrier densities for all the $Fe_{100-x}(SiO_2)_x$ samples. The calculated carrier density for pure Fe at 300 K is $16.3 \times 10^{22} / cm^3$, which is in fair agreement with the reported value $17.0 \times 10^{22} / cm^3$ [17]. This agreement indicates that the carrier density can be obtained from the OHE coefficient in the present study. As expected, with increasing the volume fraction of insulation $SiO_2$, the effective carrier density decreases monotonically. Interestingly, the carrier density decreases very sharply when x approach 30. This dramatic decrease of carrier density can only be understood with help of percolation theory [16]. Near the percolation, the electrical transport properties is governed by hoping/tunneling and quantum interference processes [16, 18].

Using the effective carrier density (in Fig. 3), AHE resistivity and longitudinal resistivity, we plotted $\ln(\rho_{xy}^{A} / n) \sim \ln(\rho_{xx})$ in Fig. 4. By fitting the data to a straight line, we obtained the value of $\gamma$ = 2.1±0.1 with the variations of nearly 3 orders in magnitude in longitudinal resistivity and more than 4 orders in AHE resistivity. The perfect scaling law of $\rho_{xy}^{A}/n \propto \rho_{xx}^{2}$ between AHE resistivity and longitudinal resistivity strongly suggests the dissipationless feature of anomalous Hall current, an evidence of the intrinsic nature of the anomalous Hall effect in iron. As far as we know, this observation is the first experiment in which the dissipationless feature of the anomalous Hall current was observed in a simple ferromagnetic metal. As pointed out in the Ref. 1, the pre-requirement for studying the dissipationless feature of the anomalous Hall current is to keep the ferromagnetic state unchanged for all the samples. In the present study, the electron transports are within Fe clusters and through hoping/tunneling between the clusters. All the iron oxide phases (FeO, $Fe_2O_3$), if exist in the samples, are quite insulating. The oxide phases may exist as Fe/$FeO_x$ core/shell clusters [19], interfaces between the Fe clusters or in the $SiO_2$ matrix, having a very weak contribution to the ferromagnetic state of the Fe clusters. Therefore, the magnetic state of Fe clusters was kept unchanged for all the samples.

One may argue that the side-jump mechanism (the extrinsic AHE model) by Berger [11] also proposed a square dependence of the AHE resistivity on the longitudinal resistivity, i.e. $\rho_{xy}^{A} \propto \rho_{xx}^{2}$. If the carrier density (n) is constant or very weakly depends on the sample composition, the behavior in Fig. 4 may also be ascribed to the side-jump mechanism. In order to clarify this issue, we plotted the AHE resistivity as a function of longitudinal resistivity in a log-log scale in Fig. 5. It is clearly seen the data cannot be fitted to a straight line, i.e., the data cannot be scaled by $\rho_{xy}^{A} \propto \rho_{xx}^{2}$ which indicates that AHE in Fe is intrinsic in nature.

A close analysis of the data in Fig.4 reveals that if we only fit the data enclosed in the circle, we may have very different exponents for the power law $\sigma_{xy}^{A}/n \propto \rho^{\gamma}$ (the inset of Fig. 4). The ranges of variation in longitudinal resistivity and AHE resistivity enclosed in the circle are comparable and even larger than some of those in the previously studies where different exponents have been obtained [4-9]. This reveals that in order to have a reliable fitting of scaling law, a wide variation in both AHE and longitudinal resistivities is essential.

The dissipationless feature of anomalous Hall current in Fe was clearly observed, which indicates that the AHE in Fe is intrinsic in nature. Our results also strongly suggest that a wide variation in both AHE and longitudinal resistivities is essential to fit a reliable scaling law.

**Acknowledgements**

The work described in this paper was supported partially by grants from the Research Grants Council of the Hong Kong Special Administrative Region (Project No. 604407) and the National Science Foundation of China (No. 50729101 and 50525101).

**Figure Captions:**

Fig. 1. Field dependent Hall resistivity obtained at different temperatures, (a) $Fe_{94.64}(SiO_2)_{5.36}$ and (b) $Fe_{59.49}(SiO_2)_{40.51}$; (c) Saturated anomalous Hall resistivity as a function of the $SiO_2$ volume fraction at 5K.

Fig. 2. (a) The temperature dependent longitudinal resistivity for samples with different $SiO_2$ volume fraction; (b) Temperature coefficients of resistivity (TCR) as a function of volume fraction of $SiO_2$.

Fig. 3. Effective carrier concentration extracted from the ordinary Hall coefficient versus the volume fraction.

Fig. 4. The normalized anomalous Hall resistivity obtained at 5K versus the longitudinal resistivity in a log-log scale for all the samples. The line is a linear fit of the data, with a slope $\gamma=2.1\pm0.1$. The inset: the zoom of the data within the circle. It shows clearly that different fittings are possible due to the narrow ranges of Hall and longitudinal resistivities.

Fig. 5. AHE resistivity (5 K) as a function of longitudinal resistivity in a log-log scale.

Fig. 1a

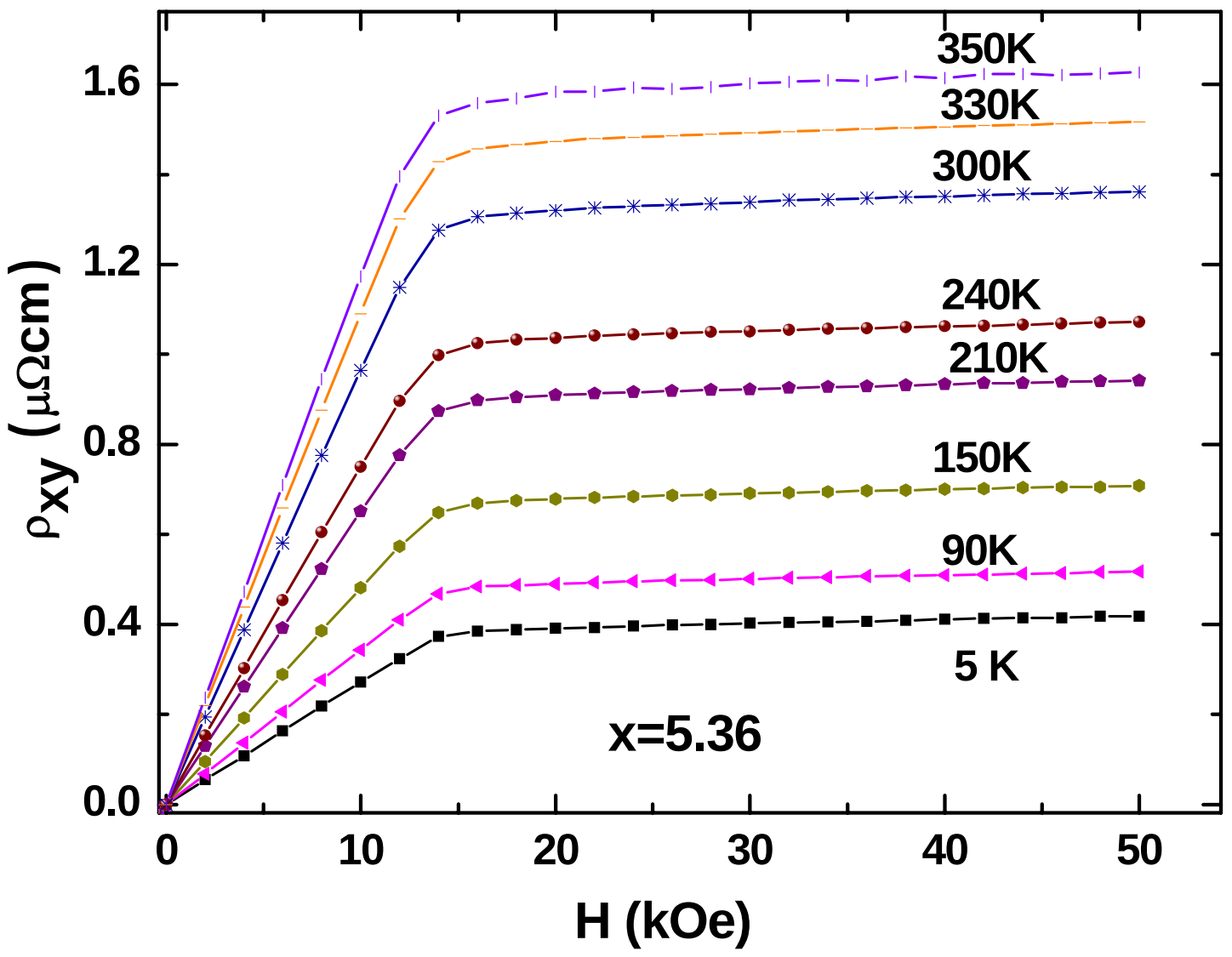

Fig. 1b

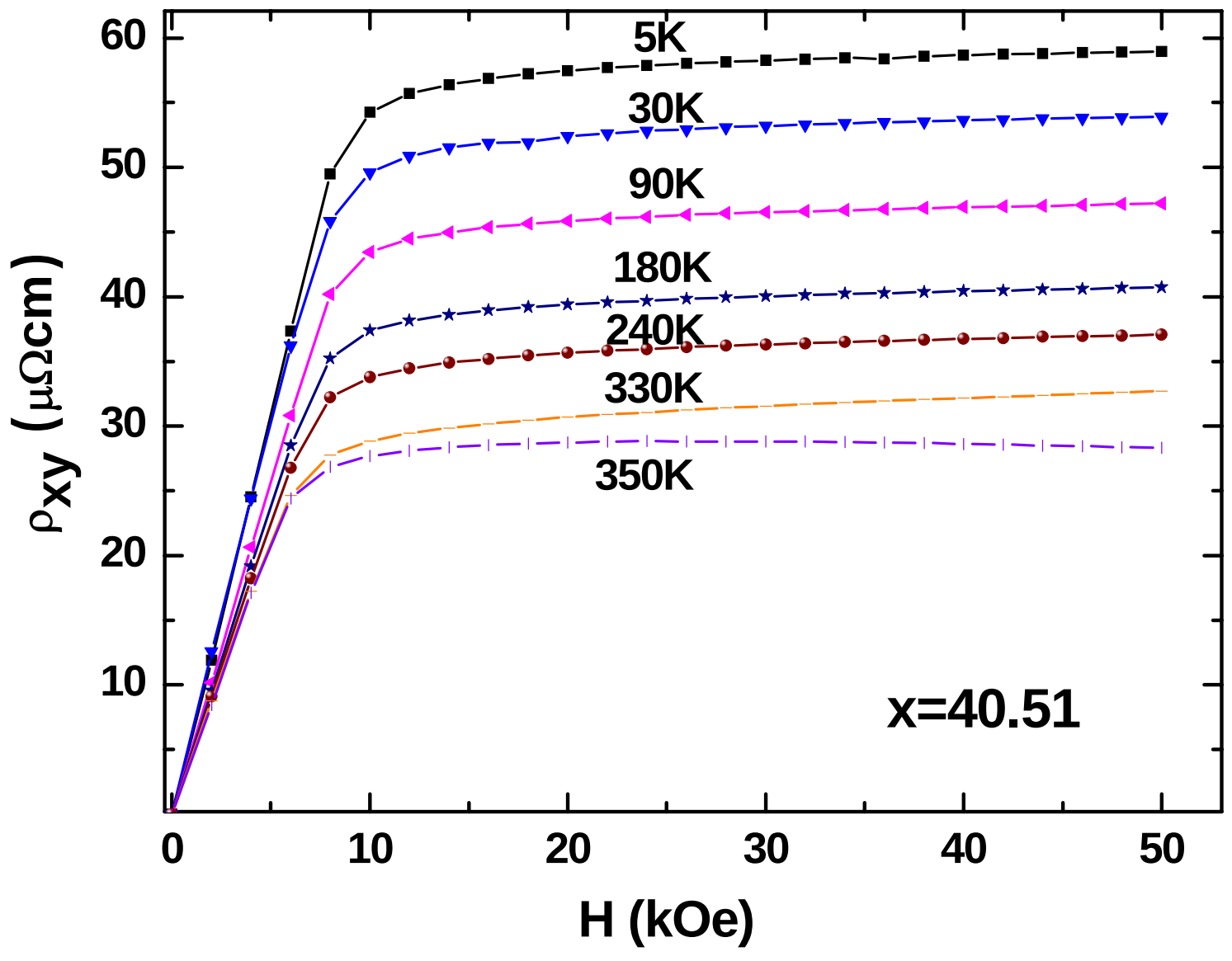

Fig. 1c

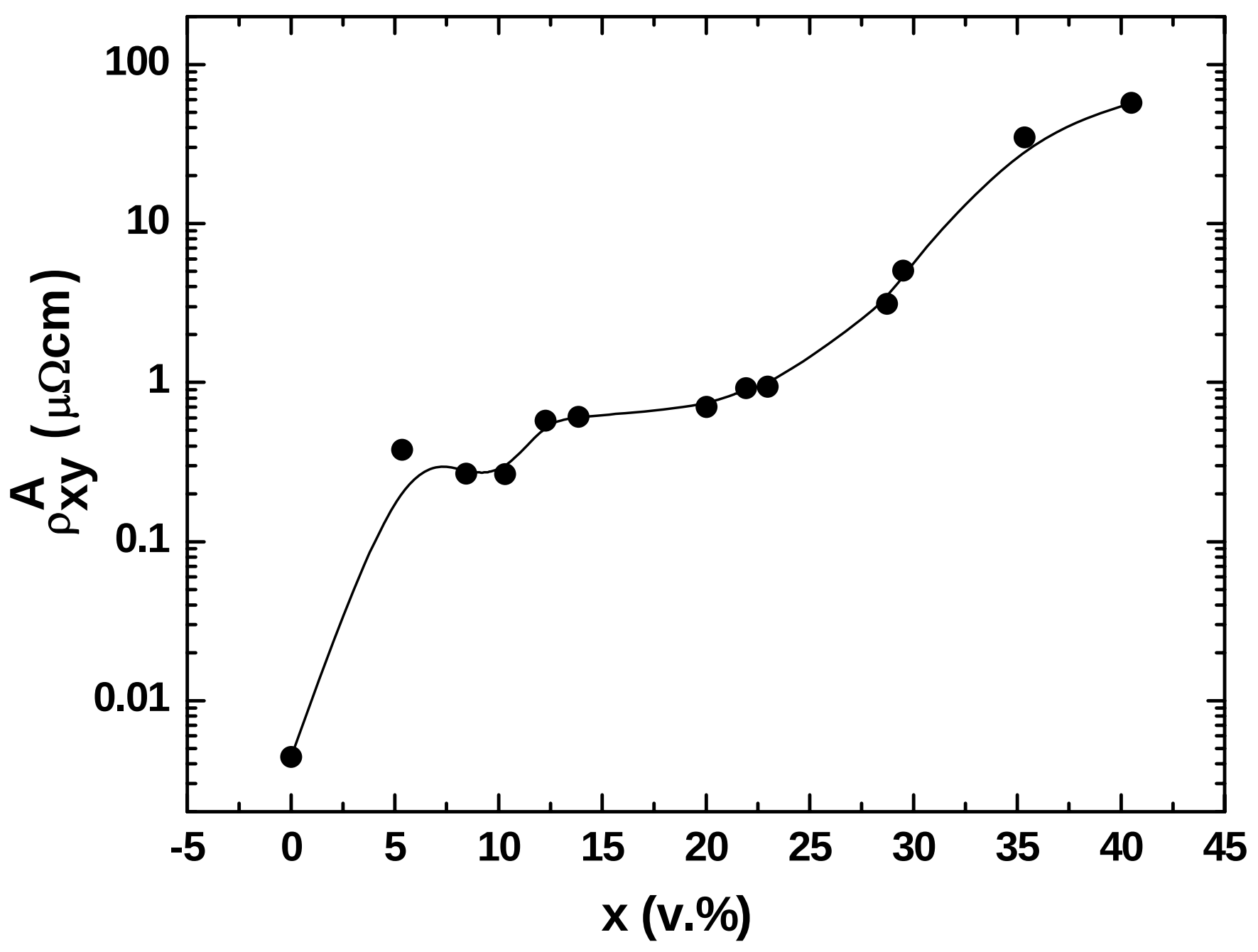

Fig. 2a

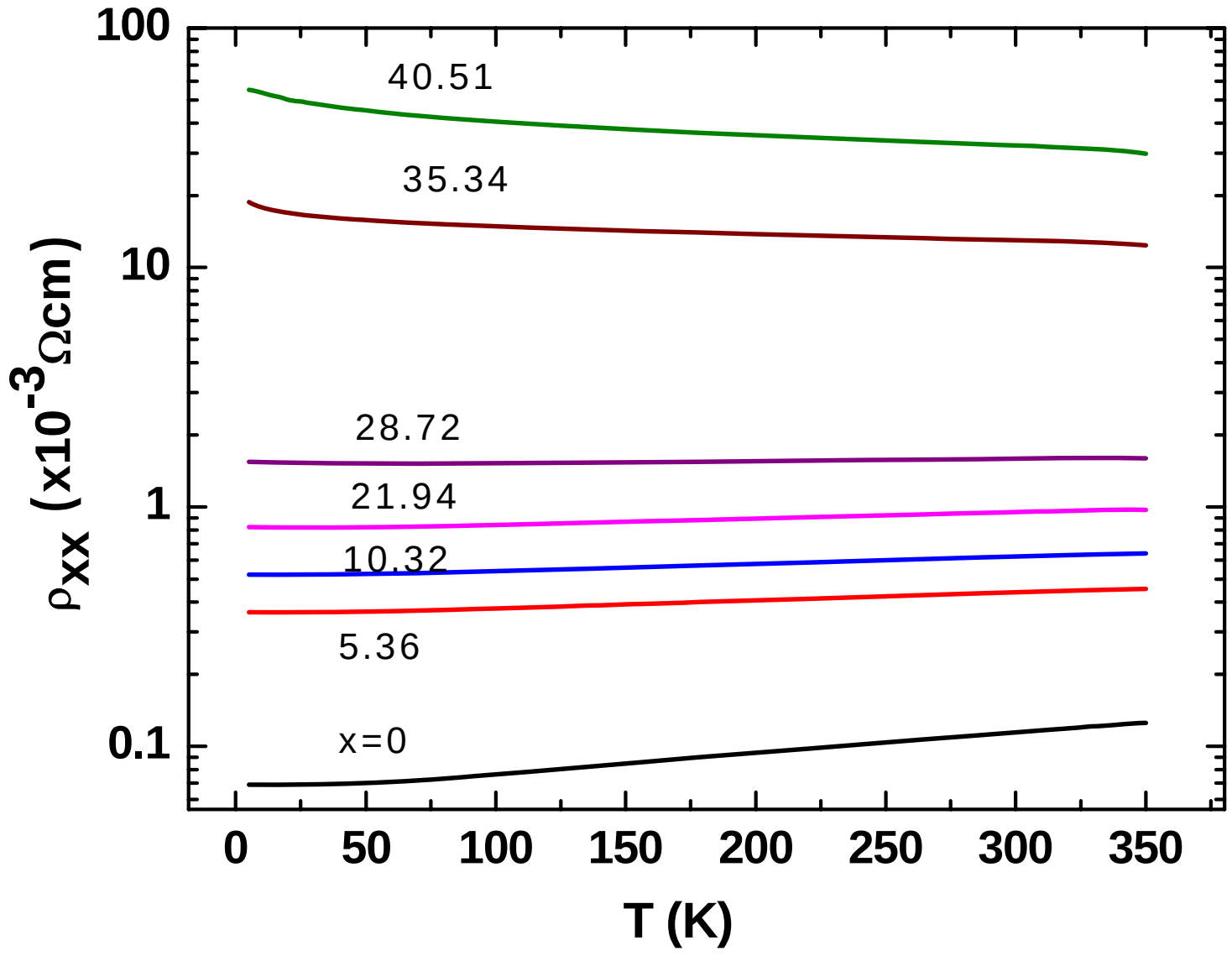

Fig. 2b

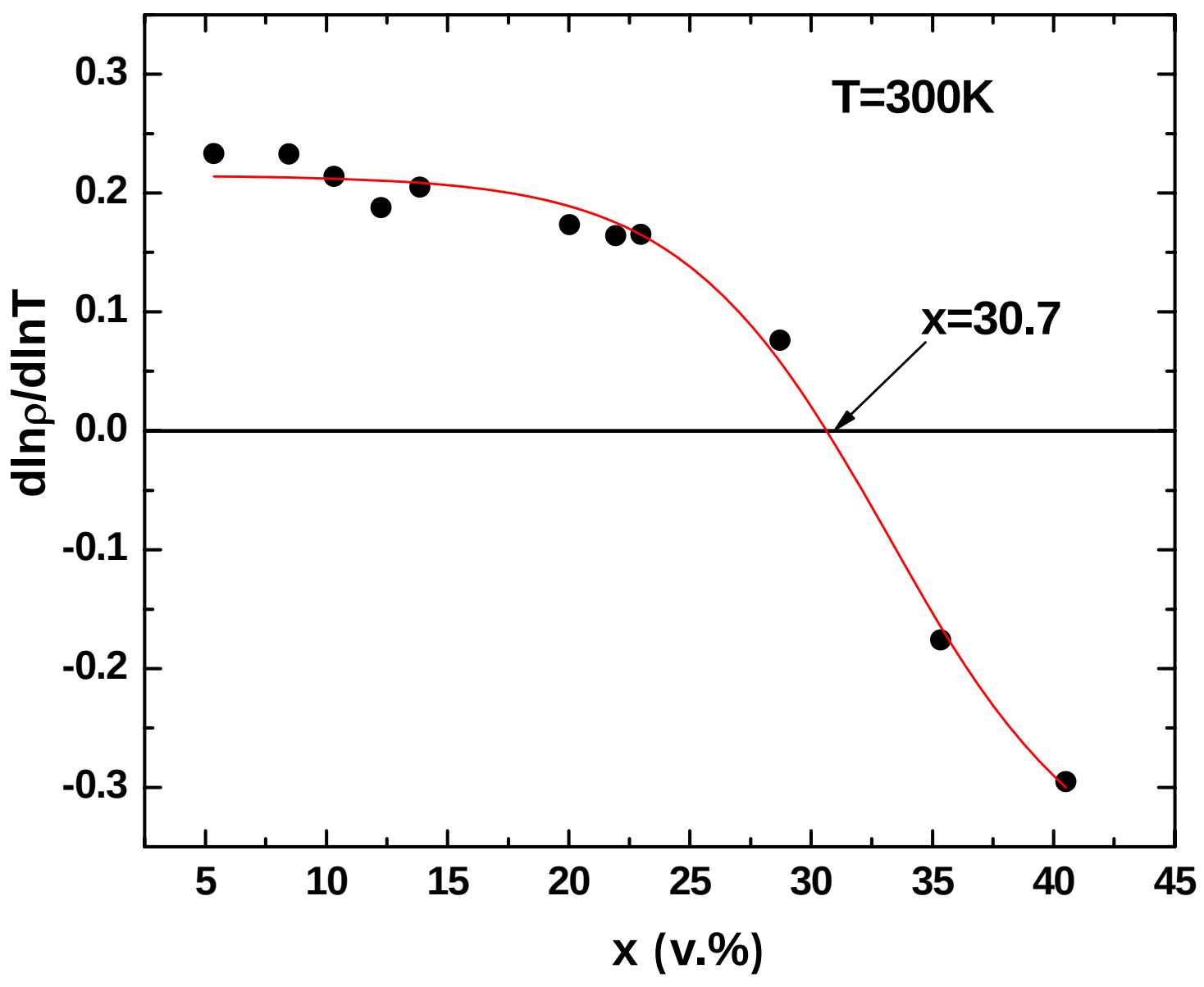

T=300K
x=30.7
dln$\rho$/dlnT
x (v.%)

Fig. 3

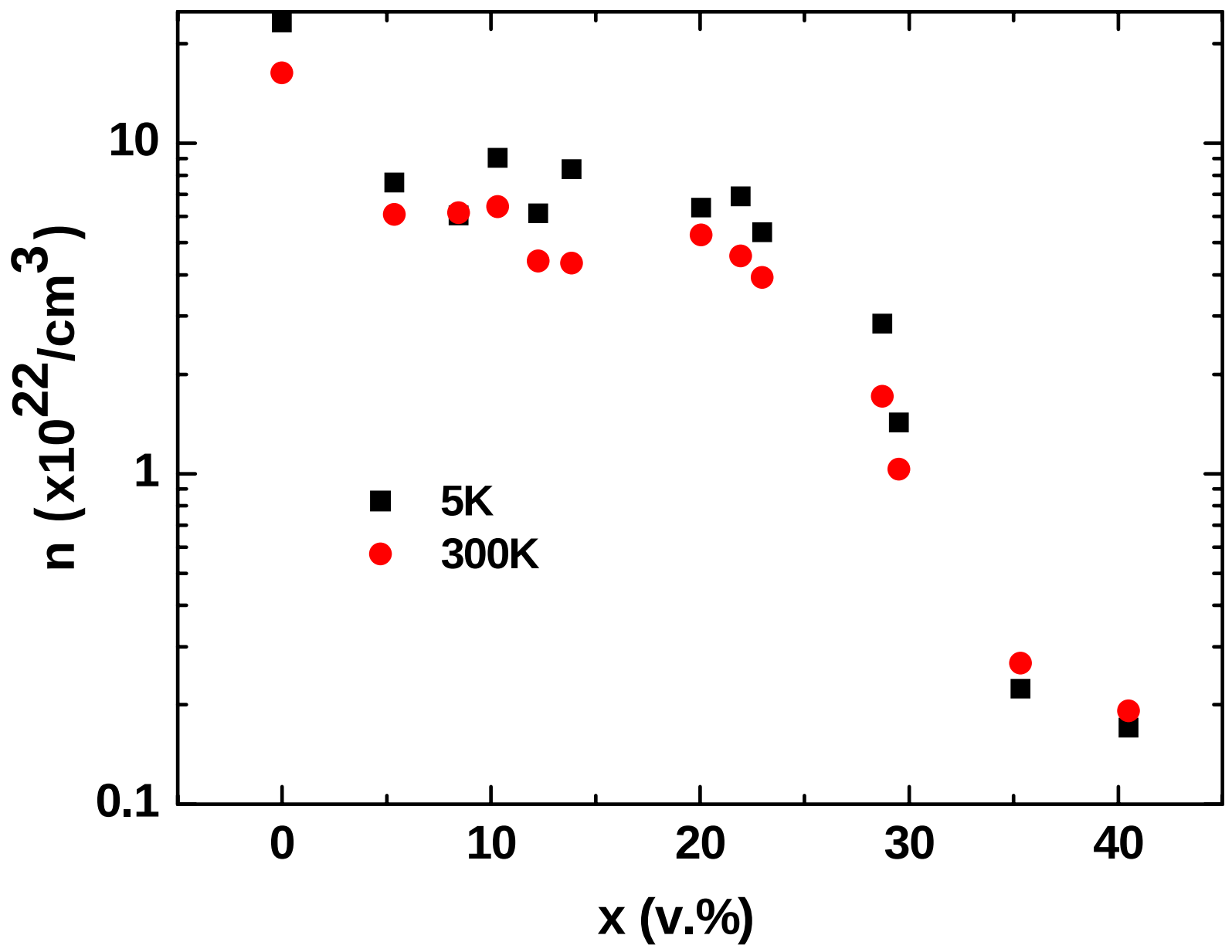

Fig.4

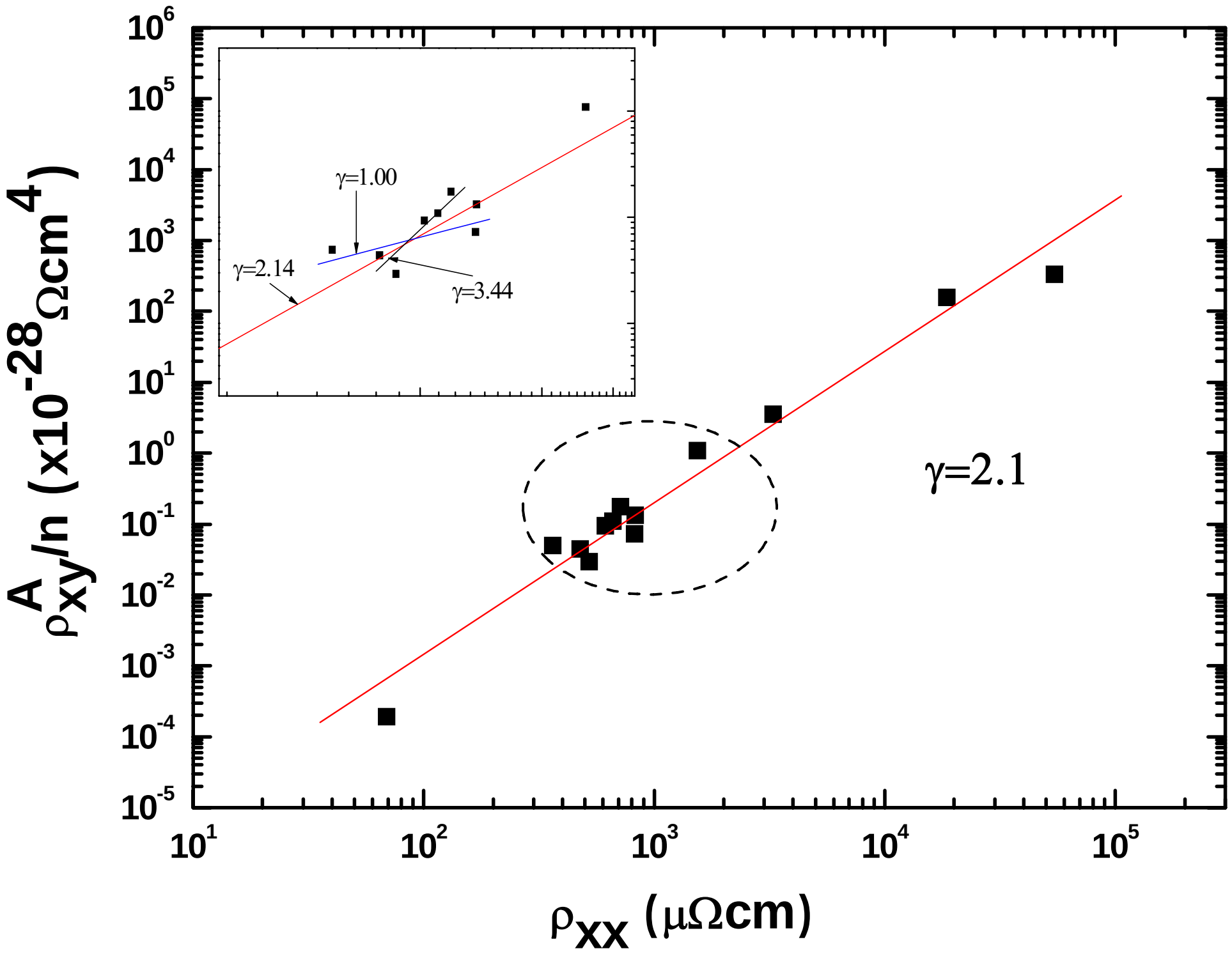

Fig.5

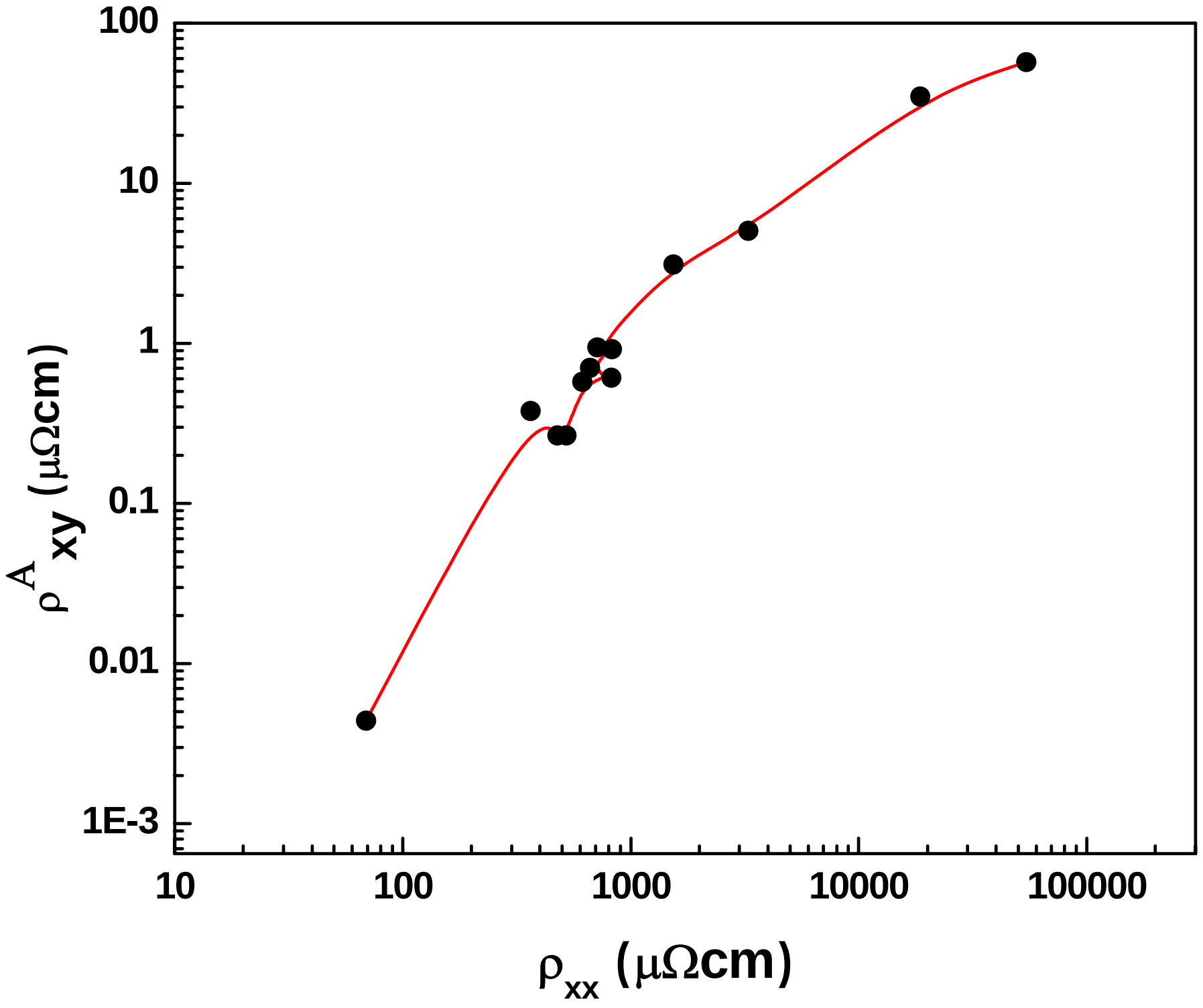